\documentclass[final,3p,times]{elsarticle}

\usepackage{amsmath}
\usepackage{amssymb}
\usepackage{bm}
\usepackage{listings}
\usepackage{color}
\usepackage{url}
\usepackage{theorem}
\usepackage{subfigure}
\usepackage{intmacros}
\usepackage{xspace}
\usepackage{main}

\journal{Journal of Computational and Applied Mathematics }

\begin{document}

\begin{frontmatter}

\title{
  Computer-Assisted Verification of Four Interval Arithmetic Operators
}

\author[a1]{Daisuke Ishii}
\ead{dsksh@acm.org}
\author[a1]{Tomohito Yabu}

\address[a1]{Department of Information Science, University of Fukui. 3-9-1 Bunkyo, Fukui-shi, Fukui, 910-8507, Japan}

\begin{abstract}

  {Interval arithmetic libraries} provide
  the four elementary arithmetic operators for operand intervals bounded by floating-point numbers. 
  %
  %
  Actual implementations need to make a large case analysis that considers, e.g.,
  magnitude relations between all pairs of argument bounds,
  positional relations between the arguments and zero, and
  handling of the special values, 
  $\pm\infty$ and $\mathrm{NaN}$.
  Their correctness is not obvious as they are implemented by human hands,  which comes to be critical for the reliability.
  This work provides a mechanically-verified interval arithmetic library.
  For this purpose, we utilize the {Why3 platform} equipped with 
  a specification language for annotated programs and back-end theorem provers.
  %
  %
  We conduct several proof tasks for each of three properties of the target code: validity, soundness, and tightness;
  \Rev{zero division exception handling is also verified for the division code.}
  \RRRev{To accomplish the proof, we propose several techniques for specification/verification.
  First, we specify additional lemmas that support deductions made by back-end SMT solvers,
  which enable to discharge proof obligations in floating-point arithmetic containing nonlinear terms.}
  \RRRev{Second, we examine the annotation of tightness, which requires to assume that a computation may result in NaN;
  we propose specific extremum operators for this purpose.}
  In the experiments, applying the techniques in conjunction with
  the Alt-Ergo SMT solver and the Coq proof assistant proved the entire code.
  %
\end{abstract}


\end{frontmatter}


\section{Introduction}
\label{s:intro}

Interval arithmetic~\cite{Moore1966,Neumaier1990} is a reliable method for numerical computation that deals with reals. It handles (machine-representable) intervals instead of numerical values, typically floating-point (FP) numbers, and evaluates arithmetic expressions via computing their interval enclosures with careful control of rounding modes and interval widening.
The four elementary arithmetic operators on intervals can be implemented in an efficient way that only considers the bounds of the given argument intervals. 
However, actual implementations need to make a large case analysis that considers, e.g.,
magnitude relations between all pairs of argument bounds,
positional relations between the arguments and zero, and
handling of the special values, 
$\pm\infty$ and $\mathrm{NaN}$.
For instance, the implementation of the multiplication in the kv library~\cite{KV} comprises \Rev{13} branches.
As existing code bases are implemented by human hands, their correctness is not obvious, which comes to be critical for the reliability.

\Rev{In this work, we use Why3 to prove the correctness of the code.}
Why3~\cite{Why3,BFMP2011} has been developed as a program verification platform, which integrates automated theorem provers (e.g. SMT solvers) and interactive proof assistants (e.g. Coq).
Why3 provides a specification language \verb|WhyML| to describe a target program and to annotate the program \Rev{with} meta-level properties. Given a specification, Why3 generates verification conditions (VCs) that ensure the correctness when they are all validated. Users can then apply the back-end provers in succession to discharge all of the VCs.
Targets of Why3 include numerical programs and several applications have been reported (e.g. \cite{Boldo2013JAR}), as support for the numerical domain has been developed both in the platform and its back-end provers.

This study aims for the verification of the four interval arithmetic operations, i.e., addition, subtraction, multiplication and division.
For this purpose, we translate an implementation into \verb|WhyML| and properly annotate the preconditions and postconditions.
For the generated VCs, we experiment with discharging them via utilizing back-end provers, \hbox{Alt-Ergo}~\cite{AltErgo} and Coq. 
Although the control structure of the implementation is simple, which only consists of branching statements, proof obligation becomes complicated due to combinations of FP number expressions and the axioms that realize the system of FP numbers/intervals in Why3.
Several VCs thus remain unproved after a basic usage of Why3.
Therefore, we examine the target annotated code and propose techniques to \Rev{complete} the verification.
Finally, \Rev{we show that the entire annotated code is verified and present experimental measures such as timings}.

\subsection{Contribution}
\label{s:intro:contr}

This \Rev{contribution is} a mechanical proof of an implementation of the four operators, which verifies mainly three properties: \emph{validity} (\RRev{``\CRRev{the} results conform to the definition of intervals''}), \emph{soundness} (``the results enclose every possible real values'') and \emph{tightness} (``the resulting enclosures are optimal/tightest'').
Our verification result indicates that the operators, which compute with the bounds of operand intervals and return the hull of computed results, satisfy the \RRev{three properties}.

\RRev{To complete} the proof, we propose several techniques to specify/verify the target interval operations and their properties.
First, \RRev{through} an examination of each unproved VC, we find that specifying an additional lemma will \Rev{help the Alt-Ergo SMT solver to discharge the VC}.
For instance, for a VC on FP numbers, a lemma on reals involving the same expression as the VC may help to discharge it.
We identify \Rev{19} lemmas needed for the complete verification.
Second, we propose a specification of the tightness property of an interval computation such that \Rev{the code annotated with} the property is verified \RRev{by} SMT solvers.
In this specification, it is essential to reformulate the interval hull operation with a consideration of 
NaN.

We note that the target of above techniques \CRRev{is} not limited to interval arithmetic operations but applicable to various proof tasks in the numerical domain,
e.g., verification of other interval functions and application programs.

\subsection{Outline}
\label{s:outline}

Section~\ref{s:ia} introduces basic notions and concepts of interval arithmetic.
Section~\ref{s:tools} explains the main tools for verification: Why3 and Alt-Ergo.
The following sections describe the main verification process for an interval arithmetic code.
First, Section~\ref{s:goal} gives an overview of the process.
Second, Section~\ref{s:code} explains the specification of the target code.
Next, Section~\ref{s:sound} and Section~\ref{s:tight} describe the processes of annotating the pre- and postconditions to the code and discharging the generated proof obligations.
Section~\ref{s:xp} reports the complete results of the verification experiment.

\section{Interval Arithmetic}
\label{s:ia}

\emph{Interval arithmetic}~\cite{Moore1966,Neumaier1990,IEEE1788} is a method for reliable numerical computations, which replaces the FP numbers that are usually used in numerical computation with intervals that enclose values in $\mathbb{R}$; the result should contain all the solutions calculated for every real in the given input interval.

Before introducing intervals, we consider sets of machine-representable numbers that conform to IEEE-754 standard~\cite{IEEE754}, i.e., \emph{floating-point (FP) numbers}.
In the following, $\mathbb{F}$ denotes a set of \emph{finite} FP numbers including the \Rev{signed} zeros ($-0$ and $+0$; \Rev{we also denote $+0$ by $0$}).
Additionally, special FP data, i.e., $-\infty$, $+\infty$, $\mathrm{NaN}$, are considered according to the context.
\Rev{
In the following, we assume IEEE-754's comparison operators $\{<, >, \leq, \geq, =\}$ for FP numbers in $\mathbb{F}\cup\{\pm\infty,\mathrm{NaN}\}$;
for example, $+0 = -0$ holds and $\mathrm{NaN} \leq x$ does not hold for $x \in \mathbb{F}\cup\{\pm\infty,\mathrm{NaN}\}$.
%
Given a real $\tilde{x} \in \mathbb{R}$, $\RndD(\tilde{x})$ and $\RndU(\tilde{x})$ denote the \emph{downward} and \emph{upward rounded values} in $\mathbb{F}\cup\{-\infty,+\infty\}$ and \Rev{are} defined by
$\RndD(\tilde{x}) := \max \{x \in \mathbb{F} \cup \{-\infty\} : x \leq \tilde{x}\}$ and 
$\RndU(\tilde{x}) := \min \{x \in \mathbb{F} \cup \{+\infty\} : x \geq \tilde{x}\}$.
Additionally, we assume the four operators for FP numbers $\odot \in \{+,-,\times,\div\}$. For $x,y \in \mathbb{F}\cup\{\pm\infty\}$, $\RndD(x \odot y)$ and $\RndU(x \odot y)$ denote the downward and upward rounded values of the operation $\odot$. Special cases are handled according to IEEE-754 and the semantics in \cite{Brain2015}: \emph{Infinity arithmetic}, e.g., $0 - +\infty = -\infty$; \emph{invalid operations}, e.g., $+\infty - +\infty = \mathrm{NaN}$; and \emph{overflows}, e.g., positive overflow of $\RndD(x \odot y)$ evaluates to $\max \mathbb{F};$
we also assume that \emph{underflows} result in correctly rounded values.
}

An \emph{interval} $\x = [\LB{x},\UB{x}]$ is defined as a closed connected set $\{\tilde{x} \in \mathbb{R} : \LB{x} \!\le\! \tilde{x} \!\le\! \UB{x}\}$, where $\LB{x} \le \UB{x}$.
In an actual implementation, the \emph{lower} and \emph{upper bounds} $\LB{x}$ and $\UB{x}$ are restricted to FP numbers 
\Rev{in $\mathbb{F}$.}
We also consider the \emph{entire interval} $[-\infty,+\infty] := \mathbb{R}$ and \emph{half-bounded intervals} $[\LB{x},+\infty] := \{\tilde{x} \in \mathbb{R} : \LB{x} \!\le\! \tilde{x}\}$ and $[-\infty,\UB{x}] := \{\tilde{x} \in \mathbb{R} : \tilde{x} \!\le\! \UB{x}\}$, where $\LB{x},\UB{x} \in \mathbb{F}$.
$\mathbb{IF}$ denotes the set of intervals, which contains the entire interval and the closed/half-bounded intervals bounded by FP numbers in $\mathbb{F} \cup \{\pm\infty\}$.
%
%
Intervals are considered as \emph{overapproximations} of (
non-empty) sets $S \subseteq \mathbb{R}$ of interest;
when the extrema of a set are not representable in $\mathbb{IF}$, we consider an interval whose lower/upper bounds are \emph{rounded} downwards/upwards.
\Rev{Given a set $S \subseteq \mathbb{R}$,
An optimal (tightest) overapproximation is obtained by the \emph{hull} operation $\mathrm{hull}(S) := \AB [\RndD(\mathrm{inf}\ S), \RndU(\mathrm{sup}\ S)]$.}
%

\RRRev{Recently, interval arithmetic has been standardized~\cite{IEEE1788}.
The standard is layered into four levels as in IEEE-754 and it is parameterized with the notions such as \emph{flavors} and \emph{decorations} to be adapted to various models and implementations.
The interval arithmetic in this paper deals with a variant of the set-based flavor.
It is different in that we consider neither empty intervals nor division by an interval containing zero. The set-based flavor represents the empty set as empty intervals, e.g. $[\LB{x},\UB{x}]$ where $\LB{x}>\UB{x}$ and $[-\infty,-\infty]$, but our $\mathbb{IF}$ does not contain them.
In the set-based flavor, $\x \div [0,0]$ is computed as an empty interval, and other divisions by intervals containing zero can be done with the two-output division scheme, whereas our division operator throws an exception.
Otherwise, our interval arithmetic is ``classical''~\cite{IEEE1788}, which is based on the ``simple model''~\cite{Pryce2006a}.
In this paper, we intend to make the interval arithmetic code simple without missing the essential verification process. We consider the process will be applied to other extensions in a future work.}

Now, we introduce a definition 
of the four \Rev{arithmetic} operations \Rev{on} intervals.
\begin{definition}[Four interval arithmetic operators]
    For $\x,\y \in \mathbb{IF}$ and $\odot \in \{ +, -, \times, \div \}$,
    \begin{equation*}
      \x \odot \y ~:=~ \mathrm{hull}(\{ x \odot y \,:\, x \in \x, y \in \y \})
        \qquad \text{(We assume $0 \not\in \y$ when $\odot = \div$)}.
    \end{equation*}
\end{definition}

%
The following theorem gives 
\RRev{the basic principle for the tight implementation of the interval operators},
namely to take the hull of the values computed with the bounds of operand intervals.

\begin{theorem}
    \label{th:impl}
    \Rev{For $\x,\y \in \mathbb{IF}$ and $\odot \in \{ +, -, \times, \div \}$
    (we assume $0 \not\in \y$ when $\odot = \div$),
    \begin{equation} \label{eq:th}
        \x \odot \y ~\RRev{\supseteq}~
        \bigl[ \MyMin \bigl\{
            \RndD(\LB{x}\odot\LB{y}),
            \RndD(\LB{x}\odot\UB{y}),
            \RndD(\UB{x}\odot\LB{y}),
            \RndD(\UB{x}\odot\UB{y})\bigr\},~
        \MyMax \bigl\{
            \RndU(\LB{x}\odot\LB{y}),
            \RndU(\LB{x}\odot\UB{y}),
            \RndU(\UB{x}\odot\LB{y}),
            \RndU(\UB{x}\odot\UB{y})\bigr\}
        \bigr],
    \end{equation}
    where for $S \subseteq \mathbb{F} \cup \{\pm\infty,\mathrm{NaN}\}$,
    \begin{align} \label{eq:minmax4}
        \MyMin\ S &~:=~ 
        \begin{cases}
            \min (S - \{\mathrm{NaN}\}) & \text{if $S \neq \{\mathrm{NaN}\}$}, \\
            0 & \text{if $S = \{\mathrm{NaN}\}$},
        \end{cases} &
        \MyMax\ S &~:=~ 
        \begin{cases}
            \max (S - \{\mathrm{NaN}\}) & \text{if $S \neq \{\mathrm{NaN}\}$}, \\
            0 & \text{if $S = \{\mathrm{NaN}\}$}.
        \end{cases}
    \end{align} }
\end{theorem}


\RRev{See \ref{a:proof} for the proof.}
\RRev{Because the above theorem serves as a basis for the tightness specification in the following sections, we only consider the one-way inclusion $\supseteq$ in \eqref{eq:th}.}
\RRRev{In the theorem, we use modified extremum operators $\MyMin$ and $\MyMax$
with an unusual handling of $\mathrm{NaN}$, which occurs in some FP computations, e.g., $+\infty - +\infty$ and $0 \times +\infty$.
They evaluate to zero when all the arguments are $\NaN$ (with the second cases), whereas the operators of IEEE-754 evaluate to $\NaN$.
%
The $\min$ and $\max$ operators in the right-hand sides of \eqref{eq:minmax4} select the extrema based on the ordinary comparison between FP numbers, e.g., $\min \{0,+\infty\} = 0$.%
}
For example, a subtraction and a multiplication are performed as:
\begin{align*}
[0,+\infty] - [0,+\infty] 
    &~\supseteq~ \bigl[ \MyMin \bigl\{0,-\infty,+\infty,\mathrm{NaN}\bigr\},~ \MyMax \bigl\{0,-\infty,+\infty,\mathrm{NaN}\bigr\} \bigr] \\
    &~=~ \bigl[ \min \{0,-\infty,+\infty\},~ \max \{0,-\infty,+\infty\} \bigr] 
    ~=~ [-\infty,+\infty], \\
[0,0] \times [-\infty,+\infty] 
    &~\supseteq~ \bigl[ \MyMin \bigl\{\mathrm{NaN}\},~ \MyMax \{\mathrm{NaN}\}\bigr]
    ~=~ [0,0].
\end{align*}
%
\RRev{In our context,}
the second cases of \eqref{eq:minmax4} only happen in the
multiplications of $[0,0]$ and $[-\infty,+\infty]$.

Several libraries that implement the interval four operators have been developed, 
e.g., PROFIL/BIAS,\footnote{\url{http://www.ti3.tuhh.de/keil/profil/index_e.html}}
filib++,\footnote{\url{http://www2.math.uni-wuppertal.de/wrswt/software/filib.html}}
Boost Interval Arithmetic Library,\footnote{\url{https://www.boost.org/doc/libs/1_72_0/libs/numeric/interval/doc/interval.htm}}
gaol,\footnote{\url{http://frederic.goualard.net/\#research-software}}
libieeep1788,\footnote{\url{https://github.com/nadezhin/libieeep1788}}
INTLAB,\footnote{\url{http://www.ti3.tu-harburg.de/rump/intlab/}}
GNU Octave interval package,\footnote{\url{https://octave.sourceforge.io/interval/index.html}} and
kv~\cite{KV}. 
In the actual code base, the multiplication and division operators are implemented as nested branching statements based on a careful case analysis.
The number of branches increases further \Rev{due to} other factors such as optimizations to reduce the number of rounding mode controls~\cite{Rump2016}, data-level parallelization~\cite{Goualard2008c} and multi-precision FP computation~\cite{KV}.

\section{Why3 Platform and Alt-Ergo SMT Solver}
\label{s:tools}

\Rev{The main tools used in this work are Why3 and Alt-Ergo.}

\subsection{\RRev{Overview of the Tools}}
\label{s:tools:overview}

\emph{Why3} (version 1.1.0)~\cite{Why3,BFMP2011} 
is a platform for deductive program verification, which provides:
(i) a {specification language} \verb|WhyML| to describe programs and specifications;
(ii) a \emph{verification condition} (VC) generator that applies \emph{weakest precondition calculus} to the annotated programs; and 
(iii) a proof intermediate interface that facilitates discharging 
the VCs using the back-end automated/interactive theorem provers.
Why3's back-end theorem provers include SMT solvers, e.g., Alt-Ergo~\cite{AltErgo}, CVC4 and Z3,\footnote{Alt-Ergo: \url{https://alt-ergo.ocamlpro.com/}; CVC4: \url{https://cvc4.github.io/}; Z3: \url{https://github.com/Z3Prover/z3}}
and theorem proof assistants, e.g., Coq.\footnote{\url{https://coq.inria.fr/}} 
The possible results of applying an automated prover to discharge a VC are valid, invalid, timeout (within a configured time limit), or unknown.
Why3 provides a GUI to allow users to browse a list of generated VCs.
Users assign a back-end prover to each VC to validate it.
When provers cannot validate a VC, users are able to modify the VC (e.g., split one VC into several VCs) and investigate a prover's input/output though the GUI.
%

\verb|WhyML| is an OCaml-like language to describe functional programs and to annotate programs with first-order predicate logic statements.
The language has a type system, which  provides the type \verb|real| for real numbers, types for FP numbers, user-defined record types, etc.
\verb|WhyML| also has a module system that encapsulates specification as a \verb|module|; basic modules are supplied in the standard library.
There are two implementations of 64-bit FP numbers: the type \verb|t| provided by the \verb|ieee_float.Float64| module and 
the type \verb|double| of the \verb|floating_point.DoubleFull| module.
Running examples in this paper uses the former, which has been introduced in a recent version of Why3; it is compatible with the \verb|FloatingPoint| theory of SMT-LIB and processed by dedicated solvers when the VCs are discharged with SMT solvers.
The latter realizes a theory of FP numbers based on the theory of reals and
the generated VCs are encoded with the predicates on reals.
\Rev{The \RRev{two} modules provide arithmetic operators (with rounding mode control) and comparison operators that conform to IEEE-754.}
\Rev{FP numbers are related \RRev{to} real numbers using the rounding functions and translation functions provided by either module.}

\emph{Alt-Ergo} (version~2.2.0)~\cite{AltErgo} is an SMT solver, which \Rev{shares various features with Why3}.
For example, Alt-Ergo has a trigger mechanism that allows users to guide the solver \Rev{towards the proof,} using a user-defined lemma.
In \verb|WhyML|, auxiliary lemmas can be specified with triggers and they are utilized by Alt-Ergo.
In our experiments, \Rev{with the help of several user-defined lemmas,} Alt-Ergo discharged the most VCs.

\subsection{Program Verification Process Using Why3}
\label{s:tools:process}

\Rev{A user of Why3 verifies a program through performing the following tasks.}


\Rev{
\emph{Specification of the target code, properties and lemmas}.
The user implements the target code in the \texttt{WhyML} language and
annotates the properties to the code as preconditions and postconditions.
Here, requirements for the code should be better represented as simple and generic properties; however, such properties tend to require more efforts to verify.
In our work, we investigate an intuitive yet provable representation of the tightness property (cf. Theorem~\ref{th:impl}).
Also, the user is able to annotate properties about intermediate states by inserting \texttt{assert} formulas in the middle of the code, and specify lemmas related to the properties besides the code.
These auxiliary annotations will help automated deductions performed by provers like Alt-Ergo.
}

\Rev{
\emph{Generation of VCs and proving with back-end provers}.
Once an annotated \texttt{WhyML} program is fed to the Why3 tool, a set of VCs are generated; discharging them all entails the correctness of the annotated program.
The user is able to modify the VCs by applying logical transformations; for example, s/he can split a VC into several VCs which are easier to prove.
Then, s/he applies a prover to each VC to check its validity.
In this work, we first apply SMT solvers, i.e., Alt-Ergo and Z3, and then invoke the Coq tool for manual proving.
}

\Rev{
\emph{Analysis on unprovable VCs and modification of the specification}.
When VCs are not provable, the user returns to the specification task to modify either the target code or the annotations.
Since the automated proving processes of VCs are incomplete in general, they may fail to prove VCs even for a correct specification.
In such cases, the user should make an effort in finding a specification that results in provable VCs.
Here, an interactive proof task with Coq may help to understand the proof context and to identify an additional lemma that is useful to discharge the proof goal.
In this work, we actually make such analyses; as a result, we insert assertions within the code and specify lemmas to help Alt-Ergo's proof process.
}

\section{Towards a Verified Interval Arithmetic Library}
\label{s:goal}

The goal of this study is to verify the correctness of an implementation of the four interval arithmetic operators.
The implementation code is first cloned from the kv library~\cite{KV} and then modified to conform \RRev{to} Theorem~\ref{th:impl} (see Section~\ref{s:code:m} for details).
To achieve this goal, 
\RRev{we apply the verification process described in Section~\ref{s:tools:process} to the operator code.
This section describes the properties we aim at verifying.}

When interval operators are \Rev{implemented} in a programming language (e.g. \texttt{WhyML}),
without loss of generality, we can assume that intervals are represented as \emph{pairs of FP numbers} such as $\x = (\LB{x},\UB{x})$ where $\LB{x}$ and $\UB{x}$ can be arbitrary FP numbers in $\mathbb{F}\cup\{\pm\infty,\mathrm{NaN}\}$;
we denote the set of such pairs by $\mathbb{IF}^*$ in this section.
\Rev{In the following, we denote that a pair $\x \in \mathbb{IF}^*$ is a valid interval by $\x \in \mathbb{IF}$.}
\Rev{Accordingly}, interval operations $\odot \in \{+,-,\times,\div\}$ will be \Rev{implemented} as procedures $\f_\odot : \mathbb{IF}^*\times\mathbb{IF}^* \to \mathbb{IF}^*$.

The properties are specified as \Rev{one} precondition and \Rev{four} postconditions, i.e., first-order conditions in $\mathbb{R}$, $\mathbb{F}$, \Rev{$\mathbb{IF}$ and} $\mathbb{IF}^*$.
Here, we present a high-level specification of the pre- and postconditions.
\Rev{
\begin{definition}[Pre- and postconditions]
    \label{def:conds}
    Let $\x, \y, \r \in \mathbb{IF}^*$ be pairs of FP numbers,
    $\odot \in \{+,-,\times,\div\}$ be an operation, and 
    $\f_\odot$ be an implementation of the interval operation, which is interpreted as a function $\mathbb{IF}^*\times\mathbb{IF}^* \to \mathbb{IF}^*$.
    Then, the \emph{precondition} $P_V$ and the \emph{postconditions} $Q_V$, $Q_S$, $Q_T$ and $Q_Z$ of computation $\r := \f_\odot(\x, \y)$ 
    are specified as follows:
\begin{align*}
    \text{(Validity precondition)} & &
    P_V &~~:\equiv~~ \x \in \mathbb{IF} \LAnd \y \in \mathbb{IF}, \\
    \text{(Validity postcondition)} & &
    Q_V &~~:\equiv~~
    \r \in \mathbb{IF}, \\
    \text{(Soundness postcondition)} & &
    Q_S &~~:\equiv~~
    \ForAll{\tilde{x}}{\x},\ \ForAll{\tilde{y}}{\y}, ~
    \tilde{x}\odot\tilde{y} \in 
    \r, \\
    \text{(Tightness postcondition)} & &
    Q_{T} &~~:\equiv~~
    \r ~=~ 
    \bigl[
        \MyMin \bigl\{
            \RndD(\LB{x}\odot\LB{y}),
            \RndD(\LB{x}\odot\UB{y}),
            \RndD(\UB{x}\odot\LB{y}),
            \RndD(\UB{x}\odot\UB{y})
        \bigr\}, \\
        & & & \hspace{5em}
        \MyMax \bigl\{
            \RndU(\LB{x}\odot\LB{y}),
            \RndU(\LB{x}\odot\UB{y}),
            \RndU(\UB{x}\odot\LB{y}),
            \RndU(\UB{x}\odot\UB{y})
        \bigr\}
    \bigr], \\
    \text{(Exceptional postcondition)} & &
    Q_{Z} &~~:\equiv~~ 0 \in \y.
\end{align*}
\end{definition}
}

\Rev{Note that, in the right-hand sides, each appearance of the operator $\odot$ acts as that for reals or FP numbers.}
In the definition, first, the precondition $P_V$ states that the operand pairs of FP numbers are \emph{valid} intervals, i.e., their lower (resp. upper) bounds are in $\mathbb{F} \cup \{-\infty\}$ (resp. $\mathbb{F} \cup \{+\infty\}$) and the magnitude relation holds between the lower and upper bounds.
Second, \Rev{similar to} $P_V$, the postcondition $Q_V$ states the \emph{validity} of the computation result.
Third, the postcondition $Q_S$ states the \emph{soundness}: ``The result contains every possible real operation results.''
Fourth, the postcondition $Q_T$ states the \emph{tightness}: ``The result is an optimal/tightest overapproximation of the four computations with the argument FP numbers.''
\Rev{Finally, $Q_Z$ is the postcondition for the case where the computation of $\f_\div$ throws the \emph{zero division exception}.}

\Rev{If the target code is executed, it will perform an operation and either terminate normally or throw an exception that signals zero division when $\odot = \div$.
For the case where the code terminates normally, we first verify that a result $\r$ is an interval that overapproximates the true results in $\{\tilde{x}\odot\tilde{y} ~:~ \tilde{x}\in\x, \tilde{y}\in\y \}$ ($Q_V$ and $Q_S$).
Secondly, we verify that $\r$ is an optimal overapproximation as described in Theorem~\ref{th:impl} ($Q_T$).
Here, we take into account the overflow cases as they are specified in the Why3's module.
Finally, for the exceptional case, we verify that the code correctly detects the zero division ($Q_Z$). 
\RRRev{Our result contains a proof of a variant of Theorem~\ref{th:impl}: %
\CRRev{Soundness of the right-hand side of \eqref{eq:th}.}}}

\Rev{The representation of the tightness in $Q_T$ is rather restrictive as it assumes the monotonicity of the operations.
More generic representation (e.g. 
$\ForAll{\tilde{r}}{\r}, \Exists{\tilde{x}}{\x}, \Exists{\tilde{y}}{\y}, \tilde{r} \in (\mathrm{hull}(\tilde{x})\odot\mathrm{hull}(\tilde{y}))$) 
is preferable but it will certainly make the verification process less automated and \RRev{more} difficult, which involves many interactive proof tasks. In this work, we aim at making the proof process automated and simple with a restrictive annotation.}

\section{Target Code: Interval Multiplication Procedure}
\label{s:code}

In the following sections, we consider interval multiplication as a running example.
This section explains how we specify intervals (Section~\ref{s:code:i}) and an interval procedure (Section~\ref{s:code:c}) with the \verb|WhyML| language.
\Rev{Also, Section~\ref{s:code:m} explains a difference between our code and the original code.
Additional notes on specifying other operators are described in \ref{a:other}.}

\subsection{Specification of Intervals}
\label{s:code:i}

\begin{figure}[t]
  \lstset{frame=single}
  \lstset{numbers=left}
  \begin{lstlisting}
module Interval
  use real.Real
  use ieee_float.Float64
  use Float64Ex

  type interval = { inf: Float64.t; sup: Float64.t }

  predicate real_in (a: real) (x: interval) =
    ((t'isFinite x.inf /\ t'real x.inf <= a) \/ is_minus_infinity x.inf) /\ 
    ((t'isFinite x.sup /\ t'real x.sup >= a) \/ is_plus_infinity x.sup)

  val constant zeroI : interval
    ensures{ is_zero result.inf /\ is_zero result.sup }

  lemma zeroR_in_zeroI: real_in 0. zeroI

  predicate valid (x: interval) =
    (t'isFinite x.inf \/ is_minus_infinity x.inf) /\ 
    (t'isFinite x.sup \/ is_plus_infinity x.sup) /\ x.inf .<= x.sup

  lemma valid_not_nan: forall x. valid x -> is_not_nan x.inf /\ is_not_nan x.sup
  lemma valid_zeroI: valid zeroI
end
  \end{lstlisting}
  \caption{Specification of the interval module.}
  \label{c:interval}
\end{figure}

\begin{table}[t]
    \centering
    \caption{\label{t:vocabulary} Some vocabularies 
    for FP numbers ($x, y \in \mathbb{F}\cup\{\pm\infty,\mathrm{NaN}\}$).}
    \begin{tabular}{ll|ll}
    \hline \hline
    Expression ~&~ Interpretation ~&~ Expression ~&~ Interpretation \\
    \hline
    $\texttt{is\char`_zero}\ x$ ~&~ 
        $\Leftrightarrow ~ \text{$x = -0$ or $x = +0$}$ ~&~
    %
    $\texttt{t'isFinite}\ x$ ~&~ 
        $\Leftrightarrow ~ x \in \mathbb{F}$ \\
    $\texttt{zeroF}$ ~&~ 
        $= ~ {+0}$ ~&~
    $\texttt{is\char`_plus\char`_infinity}\ x$ ~&~ 
        $\Leftrightarrow ~ x = +\infty$ \\
    $x\ \texttt{.<=}\ y$ ~&~ 
        $\Leftrightarrow ~ x \leq y$ (cf. IEEE-754)
        ~&~
    $\texttt{is\char`_infinite}\ x$ ~&~ 
        $\Leftrightarrow ~ \text{$x = -\infty$ or $x = +\infty$}$ \\
    $\texttt{mul\char`_down}\ x\ y$ ~&~
        $= ~ \RndD(x \times y)$ ~&~
    $\texttt{is\char`_not\char`_nan}\ x$ ~&~ 
        $\Leftrightarrow ~ \text{$x$ is not NaN}$ \\
    \hline
    \end{tabular}
\end{table}

We consider interval operations as procedures that handle data of the \verb|interval| type,
which is specified in a dedicated Why3 \verb|module| (Figure~\ref{c:interval}) named \verb|Interval|;
related constants, predicates and lemmas are also specified in this module.
In Figure~\ref{c:interval}, first, the modules for real numbers and FP numbers are imported at Lines 2--4; these modules provide type \verb|real|, type \verb|Float64.t| (double-precision FP number), and related types, constants, operators and predicates (some of them are summarized in Table~\ref{t:vocabulary}).
Most of them are specified by a set of \verb|WhyML| axioms, e.g., the operator \verb|.<=| is specified as follows.
\begin{lstlisting}[basicstyle=\ttfamily\normalsize]
forall x y. x .<= y  ->  ( (is_finite x /\ is_finite y) \/ 
  (is_minus_infinity x /\ is_not_nan y) \/ (is_not_nan x /\ is_plus_infinity y) )
\end{lstlisting}
The \verb|Float64Ex| module imported at Line~4 is our own module that provides auxiliary vocabularies for FP numbers (some are shown in Table~\ref{t:vocabulary}).
%
Second, the type \verb|interval| is defined at Line~6 as a record combining the lower (\verb|inf|) and upper (\verb|sup|) bounds of type \verb|Float64.t|.
Third, the predicate \verb|real_in| is defined at Lines~8--10 to state $a \in \x$ where $a \in \mathbb{R}$ and $\x \in \mathbb{IF}$.
\RRev{Fourth, the module specifies a constant interval 
\texttt{zeroI}, i.e., a point-wise interval containing 0.
It is specified using the \texttt{is\char`_zero} predicate instead of specifying a concrete interval to represent either of intervals whose bounds are $-0$ or $+0$.}
At Line~15, there is a lemma characterizing \verb|zeroI|, which may be used in an automated proof process.
In the rest of the specification (Lines~17--22), 
the predicate \verb|valid x| to state the validity 
of the value \verb|x| of type \verb|interval| is specified.
\Rev{For instance, it is used in the precondition $P_V$ to state that $\x \in \mathbb{IF}$.}
Then, two lemmas using \verb|valid| are specified;
\verb|valid_not_nan| states that the bounds of valid intervals are not $\mathrm{NaN}$;
\verb|valid_zeroI| states the validity of the constant specified above.
Lemmas in a \verb|WhyML| module are proved by Why3's back-end provers.
The three lemmas specified so far are proved using Alt-Ergo, where each proof takes less than 0.2s.

\subsection{Specification of Multiplication Procedure}
\label{s:code:c}

\begin{figure}[t]
  \lstset{frame=single}
  \lstset{numbers=left}
  \begin{lstlisting}
module IntervalMul
  use real.Real
  use ieee_float.Float64
  use Float64Ex
  use Interval

  let multiply (x y: interval) : interval
    (* pre- and postconditions are specified here *)
  = if x.inf .>= zeroF then
      if x.sup .= zeroF then
        zeroI
      else
        if y.inf .>= zeroF then
          if y.sup .= zeroF then
            zeroI
          else 
            { inf = mul_down x.inf y.inf; sup = mul_up x.sup y.sup }
        else if y.sup .<= zeroF then
          { inf = mul_down x.sup y.inf; sup = mul_up x.inf y.sup }
        else 
          { inf = mul_down x.sup y.inf; sup = mul_up x.sup y.sup }
    (* other 8 branches are omitted *)
end
  \end{lstlisting}
    \caption{Code snippet for multiplication of intervals: \Rev{Case $\LB{x} \geq 0$}.}
  \label{c:mul}
\end{figure}

Interval multiplication code written as a \verb|WhyML| function \verb|multiply| is shown in Figure~\ref{c:mul};
only five branches \Rev{are shown} due to space limitations, but the full specification consists of 13 branches.
At Lines~2--5, necessary modules are imported.
Multiplication is defined by the \verb|WhyML| function \verb|multiply| at Lines~7--22.
The procedure is specified as a nested \verb|if|-\verb|then| statement and the expressions inside are specified with predefined vocabularies for FP numbers and intervals.
%

\subsection{Modification of the Target Code}
\label{s:code:m}

\Rev{In this work, one major modification was made against the codebase of the kv library (version~0.4.38).
In the modification, Lines~11 and 15 of the code in Figure~\ref{c:mul} were simplified;
Line~11 of the original code was implemented as follows.}
\begin{lstlisting}[basicstyle=\ttfamily\normalsize,escapechar=@]
        if y.inf .= minus_inf || y.sup .= plus_inf then entireI else zeroI
\end{lstlisting}
\Rev{Line~11 corresponds to the case where $\x = [0,0]$ and the modified code results in $[0,0]$, whereas the original code checks whether a $\y$'s bound is $\pm\infty$ and returns either $[-\infty,+\infty]$ or $[0,0]$.
Line~15 is likewise for switched $\x$ and $\y$.
Thus, the code comprises 17 branches.}
\RRRev{Our verification result suggests that the original code of kv unnecessarily enlarges some results from $[0,0]$ to $[-\infty,+\infty]$, which is not correct regarding our tightness property. 
The original code is correct in terms of the cset model~\cite{Pryce2006a} but it is not clear whether this design choice was made or not (cf. the code does not support divisions by an interval containing zero).
%
Based on our interval arithmetic flavor (Section~\ref{s:ia}), our code is simplified and results in tighter intervals yet it is verified to be correct.
}

\section{Validity and Soundness of Multiplication}
\label{s:sound}

This section describes the verification process for the validity $Q_V$ and soundness $Q_S$ of the interval multiplication, in which we annotate \Rev{the target code with the properties} (Section~\ref{s:sound:a}) and discharge the resulting VCs using the Why3's back-end provers (Section~\ref{s:sound:p}).

\subsection{Specification of the Pre- and Postconditions}
\label{s:sound:a}

We annotate \Rev{the \texttt{multiply} function in Figure~\ref{c:mul} with the properties in Section~\ref{s:goal}}. 
The following pre- and postconditions, which encode $P_V$, $Q_V$ and $Q_S$ of Definition~\ref{def:conds}, respectively, are inserted at Line~8 of the code.
\footnote{The variable \texttt{result} is a built-in variable that represents the resulting value $\r$ (Definition~\ref{def:conds}) of the function under verification.}
\begin{lstlisting}[basicstyle=\ttfamily\normalsize,escapechar=@]
  requires { valid x /\ valid y } (* P_V *)
  @\vspace{-1em}@
  ensures { valid result } (* Q_V *)
  @\vspace{-1em}@
  ensures { forall u v.  real_in u x /\ real_in v y -> 
    real_in (u * v) result } (* Q_S *)
\end{lstlisting}

\subsection{Proof}
\label{s:sound:p}

Next, we generate a proof obligation (i.e., a set of VCs) for the function \verb|multiply| using the Why3 tool and try to discharge (i.e., prove the validity of) the VCs using one of the back-end SMT solvers.
This process can be easily performed through the Why3 GUI.
Here, two VCs for \CRRev{the postconditions} are generated.
However, for these VCs, \Rev{neither proof process of Alt-Ergo and Z3 does terminate after 10min}.
For that case, we can \emph{split} the VC into \Todo{26 partial VCs} (again through Why3's GUI), each of which corresponds to one of the \Todo{13 branches} and one of the two postconditions;
discharging them all entails the validity of the original VC.
We apply Alt-Ergo to prove the split VCs; it discharges \Todo{13} of them and leaves \Todo{13} VCs unproved.

To prove the \Rev{remaining} VCs, we \Rev{try} to prove them interactively using the Coq proof assistant.
\Rev{An interactive proof is pessimistic due to complication of the VCs involving intervals and FP numbers, which requires a number of rewritings to accomplish the proof.%
\footnote{\RRev{In a Coq proof context, FP numbers are formalized by real numbers with additional data; vocabularies for intervals and FP numbers are defined and characterized based on reals and integers. For instance, rewriting a context by applying a lemma on real numbers requires to unfold the definitions of the vocabularies used in the context, which \CRRev{additionally} requires a number of rewritings.}}
The proof of such VCs are performed by SMT solvers more efficiently by automating the rewritings when their search strategies are sufficient.} However, this interaction is useful to identify an auxiliary lemma that is helpful for SMT solvers.
Indeed, for the VCs under consideration, we find that the following lemma is useful for the proof with Alt-Ergo.
\begin{lstlisting}[basicstyle=\ttfamily\normalsize]
  lemma Rmult_le_compat: forall r1 r2 r3 r4 : real.
    0. <= r1 <= r2 /\ 0. <= r3 <= r4 -> r1 * r3 <= r2 * r4
\end{lstlisting}
By adding this lemma within the same module, two of the unproved VCs\footnote{Each unproved VC corresponds to the third branch shown in Figure~\ref{c:mul} and either $Q_V$ or $Q_S$, respectively.} are discharged using Alt-Ergo.
\Rev{The above lemma assumes the branching condition in Figure~\ref{c:mul} (in reals) and describes a magnitude relation between the products of the pairs of reals. The same lemma is considered in the proof in \cite{AM2010}.}
\Rev{The lemma will be regarded as a relation between the branching condition and the resulting state, when the variables are converted to the \texttt{Float64.t} type.}
The lemma is difficult to prove with SMT solvers because of the nonlinear terms but it is proved with Coq.\footnote{It is simply proved by the tactic ``\texttt{auto with real.}''}
\Rev{In this way, a proof task involving nonlinear term is done manually.}
Likewise, all of the unproved VCs \Rev{are} discharged with four additional lemmas and finally the soundness of \verb|multiply| is proved.
%
%
Overall, Alt-Ergo takes \RRev{243}s to discharge the 26 VCs. \Rev{Z3 is able to discharge four of them.}

\section{Tightness of Multiplication}
\label{s:tight}

As a second round of the verification, we verify the tightness $Q_T$ of the \verb|multiply| function.
The postcondition states that results are the hull of the candidate bounds, \Rev{i.e., rounded values of $\LB{x}\times\LB{y}, \LB{x}\times\UB{y}, \UB{x}\times\LB{y}$ and $\UB{x}\times\UB{y}$}.
However, annotation of $Q_T$ is not straightforward due to the complicated comparison of FP numbers that can be the special value NaN.
This section describes
comparison functions for our purpose (Section~\ref{s:tight:comp}), 
how we annotate $Q_T$ into the code in Figure~\ref{c:mul} (Section~\ref{s:tight:a}) and a proof process (Section~\ref{s:tight:p}).

\subsection{Comparison of Four FP Numbers}
\label{s:tight:comp}

\Rev{To} specify the postcondition $Q_T$, 
\Rev{we introduce the functions \texttt{min4} and \texttt{max4}, which correspond to (the first cases of) the $\MyMin$ and $\MyMax$ operations in Theorem~\ref{th:impl};
these functions take the four candidate bounds and \RRev{evaluate} to the minimum/maximum of them while ignoring $\mathrm{NaN}$}; %
\CRRev{they evaluate to $\mathrm{NaN}$ when all the arguments are $\mathrm{NaN}$.}
Specification of \verb|min4| is shown in Figure~\ref{c:min4}, which is actually described in the \verb|Float64Ex| module;
\verb|max4| is specified in the same way.
\Rev{First, the \texttt{min2} function that compares two FP numbers is specified (\CRRev{Lines}~1--2). \texttt{min2} ignores $\mathrm{NaN}$ given as an argument so that the expression \texttt{min2 x y} evaluates to \texttt{x} when \texttt{y} evaluates to $\mathrm{NaN}$.
Second, the \texttt{min4} function is simply specified with \texttt{min2} (Line~4).  }
%
\Rev{To support deductions when \texttt{min4} is used, we also specify six lemmas including \texttt{min4\char`_feq} (Lines~6--8), \texttt{min4\char`_fle} (Lines~10--12) and \texttt{min4\char`_feq\char`_w} (Lines~14--16) in Figure~\ref{c:min4}.
The first two lemmas state the relations between the value of \texttt{min4} and the arguments that are not $\mathrm{NaN}$;
they are respectively proved by Alt-Ergo in \RRev{7s and 0.5s}.
The last four lemmas state possible resulting values of \texttt{min4} expressions, which are equivalent to one of the arguments \texttt{w}, \texttt{x}, \texttt{y} \CRRev{and} \texttt{z};
these four lemmas are respectively proved by Alt-Ergo in 0.4s.}
\RRRev{The descriptions ``\texttt{[min4 w x y z]}'' specify the trigger of a lemma, which enforces to assume the lemma when a \texttt{min4} predicate becomes true.}

\begin{figure}[t]
  \lstset{frame=single}
  \lstset{numbers=left}
  \begin{lstlisting}
  function min2 (x y: Float64.t) : Float64.t
  = if x .< y then x else if is_not_nan y then y else x

  function min4 (w x y z: Float64.t) : Float64.t = min2 w (min2 x (min2 y z))

  lemma min4_feq: forall w x y z [min4 w x y z].
    is_not_nan w \/ is_not_nan x \/ is_not_nan y \/ is_not_nan z -> 
      min4 x y z w .= w \/ min4 x y z w .= x \/ min4 x y z w .= y \/ min4 x y z w .= z

  lemma min4_fle: forall w x y z [min4 w x y z].
    (is_not_nan w -> min4 w x y z .<= w) /\ (is_not_nan x -> min4 w x y z .<= x) /\ 
    (is_not_nan y -> min4 w x y z .<= y) /\ (is_not_nan z -> min4 w x y z .<= z)

  lemma min4_feq_w: forall w x y z [min4 w x y z].
    is_not_nan w /\ (w .<= x \/ is_nan x) /\ (w .<= y \/ is_nan y) /\ (w .<= z \/ is_nan z) -> 
      min4 w x y z .= w

  (* lemmas min4_feq_(x|y|z) are omitted *)
  \end{lstlisting}
  \caption{Specification of the \Rev{\texttt{min2} and \texttt{min4} functions}.}
  \label{c:min4}
\end{figure}

\subsection{Specification of Tightness}
\label{s:tight:a}

Before annotating $Q_T$, we specify \Rev{the conditions for the second cases of Equation~\eqref{eq:minmax4} in Theorem~\ref{th:impl}, which hold when the four candidate bounds become NaN; \CRRev{these cases} only \CRRev{happen} in multiplication.}
The condition for $\MyMin$ is specified as a predicate in the \verb|IntervalMul| module as follows (the condition for $\MyMax$ is specified likewise as \verb|mul_nan_case_sup|).
%
%
\begin{lstlisting}[basicstyle=\ttfamily\normalsize]
  predicate mul_nan_case_inf (x y: interval) =
    is_nan (mul_down x.inf y.inf) /\ is_nan (mul_down x.inf y.sup) /\
    is_nan (mul_down x.sup y.inf) /\ is_nan (mul_down x.sup y.sup)
\end{lstlisting}

Then, the pre- and postconditions $P_V$ and $Q_T$ are specified as follows.
\begin{lstlisting}[basicstyle=\ttfamily\normalsize,escapechar=@]
  requires { valid x /\ valid y } (* P_V *)
  @\vspace{-.7em}@
  ensures { not mul_nan_case_inf x y -> 
    result.inf .= min4
      (mul_down x.inf y.inf) (mul_down x.inf y.sup) 
      (mul_down x.sup y.inf) (mul_down x.sup y.sup)} (* Q_T lb *)
  @\vspace{-.7em}@
  ensures { not mul_nan_case_sup x y -> 
    result.sup .= max4
      (mul_up x.inf y.inf) (mul_up x.inf y.sup) 
      (mul_up x.sup y.inf) (mul_up x.sup y.sup) } (* Q_T ub *)
  @\vspace{-.7em}@
  ensures { mul_nan_case_inf x y /\ mul_nan_case_sup x y -> 
    is_zero result.inf /\ is_zero result.sup } (* Q_T NaN case *)
\end{lstlisting}
Here, $Q_T$ is specified in \Rev{three parts}.
\Rev{The first and second parts describe the first cases of Equation~\eqref{eq:minmax4} using the \texttt{min4} and \texttt{max4} functions, 
each of which specifies the lower or upper bound portion of $Q_T$.
The third part describes the second cases of Equation~\eqref{eq:minmax4} 
in conjunction because they only happen at the same time.}

\subsection{Proof}
\label{s:tight:p}

The first and second postconditions are transformed into 26~VCs (with three split operations) and Alt-Ergo discharges \RRev{22} of them; \RRev{4}~VCs are left unproved after 10min.
The third postcondition 
is transformed into a VC and discharged by Alt-Ergo in \Rev{0.5}s.

After investigation \Rev{of the remaining VCs}, we have found that the following lemma helps Alt-Ergo to prove the VCs generated from the \Rev{first} postcondition.
\begin{lstlisting}[basicstyle=\ttfamily\normalsize]
  lemma mul_down_positive_finite: forall x y [mul_down x y].
    is_positive (mul_down x y) /\ t'isFinite (mul_down x y) -> 
      round RTN (t'real zeroF) <= round RTN (t'real x * t'real y)
\end{lstlisting}
The lemma details the relation between the lower bound of the result and zero.
The premise extracts the unproved two branches, each of which corresponds to a remaining VC;
for example, one of the VCs corresponds to the third branch in Figure~\ref{c:mul}, where $\texttt{x.inf} \geq 0$ and $\texttt{y.inf} \geq 0$ hold and the lower bound of the result is computed as \verb|mul_down x.inf y.inf|.
We presume that applying \verb|round| explicitly on both sides of the consequent inequality facilitates unifying it with expressions specified in the \verb|ieee_float| module and thus further deductions are invoked.
Another lemma \verb|mul_up_negative_finite| is specified likewise for the \Rev{second} postcondition.
In our experiment with Alt-Ergo, 
the two lemmas are proved in \Rev{0.2}s and the 26 VCs generated from the target code are proved in \RRev{308}s.

\section{Implementation and Experiments}
\label{s:xp}

\Rev{This section summarizes how the four operators have \RRev{been} specified (Section~\ref{s:xp:spec}) and reports the verification result (Section~\ref{s:xp:data}).
Section~\ref{s:xp:extract} describes how we extracted an executable code from the specification.}
\Rev{The} file containing the specifications \Rev{and the extracted code} is available at {\url{https://dsksh.github.io/vint-jcam2020.zip}}.

\subsection{Specification}
\label{s:xp:spec}

Each operator was implemented as a \verb|WhyML| function 
\Rev{(see \ref{a:other} for some details).}
The number of branches in each function is shown in the second column of Table~\ref{t:result}.
\Rev{Each function was annotated with the conditions $P_V$, $Q_V$, $Q_S$ and $Q_T$ (the code for division was also annotated with $Q_Z$).}
Since the ``NaN case'' described in Section~\ref{s:tight:a} only happened in multiplication, $Q_T$ was \Rev{translated} as two postconditions (``\verb|Q_T lb|'' and ``\verb|Q_T ub|'' without preconditioning) for \Rev{the} other three operators.
In the experiment, we identified that additional lemmas were necessary for the verification (the numbers \Rev{of such lemmas} are shown in \Rev{the third column of} Table~\ref{t:result}).
First, two variations of the lemma in Section~\ref{s:tight:p} were needed to verify $Q_T$ for each operator code.
Second, the subtraction code was annotated with an \verb|assert| formula to detail a property of the resulting value.
Third, the multiplication and division codes were also annotated with four and seven lemmas, respectively, which were the variations of the lemma in Section~\ref{s:sound:p};
\Rev{here, the seven lemmas for division include the four lemmas for multiplication. There were 19~lemmas specified in total.}
Other than the operator code, two \verb|WhyML| modules, \verb|Float64Ex| and \verb|Interval| were prepared, which contained \Rev{30} and three lemmas \Rev{about} FP numbers and intervals, respectively.
%

\subsection{Statistical Data}
\label{s:xp:data}

\Rev{We performed an experiment to verify the implementation of four operators and all of them were successfully verified.}
The experiments (including the running examples) were operated using a 2.2GHz Intel Xeon E5-2650v4 processor with 128GB of RAM.
The following tools were used:
Why3~1.1.0, Alt-Ergo~2.2.0, Coq~8.8.0 and Z3~4.7.1.
The memory used was limited to 4GB and the time to 10min for each VC proof.

Why3's split strategy allows splitting VCs in several degrees.
Therefore, the verification was performed with three settings:
Without using the split strategy (``No split'');
with splitting VCs as many times as possible (``Split full'');
with an optimal number of splittings that requires minimum CPU time  (``Split optimal'').
Note that splitting a VC does not always result in VCs that are discharged by SMT solvers more efficiently than the original.

\begin{table}[t]
    \centering
    \caption{\label{t:result} \Rev{Experimental results.}}
    \begin{tabular}{c|c|c|cr|cr|cr} 
        \hline \hline
        & & & \multicolumn{2}{|c|}{No split} & \multicolumn{2}{|c|}{Split full} & \multicolumn{2}{|c}{Split optimal} \\
        Op. & \# Br's & \# Lm's & \# VCs & Time & \# VCs & Time & \# VCs & Time \\
        \hline
        $+$ & 1       & 2   & 3 & 36.4s  & 8   & 25.0s  & 6  & 24.7s \\
        $-$ & 1       & 2   & 3 & 63.0s  & 10  & 47.2s  & 8  & 47.0s \\
        $\times$ & 13 & 2+4 & 3 & 140s & 100 & 508s & 3  & 140s \\
        $\div$   & 7  & 6+7 & 7 & TO (1) & 33  & TO (4) & 15 & 578s \\
        \hline
    \end{tabular}
\end{table}

Herein, we present the experimental results in Table~\ref{t:result}. 
The first three columns represent
the target operator (``Op.''), the number of branches (``\# Br's'') and the number of additional lemmas (``\# Lm's'').
The data ``$m$+$n$'' indicates that $m$ lemmas were discharged with SMT solvers and $n$ lemmas were proved with Coq (only the VCs generated from the former are taken into account in the following columns).
The following columns are separated for each of the settings described above;
for each section, the number of generated VCs (``\# VCs'') and the total execution time by Alt-Ergo (``Time'') are shown.
``TO ($n$)'' \Rev{(for ``Time Out'') means} that the proofs for $n$ VCs did not finish within 10min.
\Rev{Note that the verification of the division code succeeded only with the ``Split optimal'' setting.}

Additionally, \Rev{33}~lemmas in the \verb|Float64Ex| and \verb|Interval| modules were discharged by Alt-Ergo in \Rev{32.6}s in total.
We also tried to discharge the ``Split full'' VCs with Z3, which was less efficient in our experiment;
for each operator, \Rev{1/8, 1/10, 48/100 and 2/33 VCs} were proved, respectively, and the rest resulted in timeout.
%

\subsection{Extraction of Executable Code}
\label{s:xp:extract}

\Rev{Why3 provides an \texttt{extract} function that generates an executable code from a \texttt{WhyML} code.
Using this function, we extracted an OCaml code that implements the interval type and the four operators.
The extraction of numerical programs with Why3 required additional implementations. }
\Rev{First, we prepared an OCaml module for FP arithmetic operations with rounding mode control. We implemented the \CRRev{rounding} operators in C, which were then interfaced with OCaml functions of the dedicated \texttt{Round} module. }
\Rev{Second, we extended Why3's \texttt{extract} function to support the \texttt{Float64.t} type and related constants/functions. For this purpose, we prepared a driver file that defines various translation rules. }

\section{Related Work}
\label{s:related}

Ayad and March\'{e}~\cite{AM2010} reported a case study on verifying an interval multiplication code.
They used Frama-C, which is a verification platform like Why3 equipped with SMT solvers, and verified the soundness $Q_S$ of a variant of our code.
In this paper, we go further by considering the tightness $Q_T$.

For generic FP computation, a number of verified programs have been reported.
A notable implementation is CRlibm~\cite{CRlibm}, which \Rev{aims at replacing} the mathematical library of C.
Elementary functions of CRlibm were formally verified yet \Rev{preserve} the efficiency of the computation.
The correctness was proved by analyzing the bounds of numerical errors by hand or using the Gappa tool~\cite{Daumas2010}. 
CRlibm differs from our verified code in two aspects:
First, the properties and the target programs are different; in CRlibm, it is verified that computation results are rounded values in the prescribed direction for the elementary functions such as logarithm functions.
Second, CRlibm is implemented in C and executable but our code is not; \RRRev{instead, we extract an executable OCaml code (Section~\ref{s:xp:extract}).}

Verification results for other numerical programs/algorithms have been reported~\cite{Boldo2011a,Boldo2013JAR,Roux2016}, which used tools including Frama-C, SMT solvers, Gappa and Coq (with Flocq~\cite{Boldo2011}).
They targeted computations of certain expressions in reals, which differed from our subject: operations on FP intervals.
In the verification, they aimed \Rev{at certifying} rounding error bounds of numerical computations,
as in the verification of CRlibm.
On the other hand, our target code outputs intervals that should enclose possible errors;
then, we verify the correctness of the code.
\Rev{The verification tool Frama-C is able to target C++ programs (with the C++ plugin).
Verification of our target program with Frama-C will be a future work as different tools might require different efforts to accomplish a verification.}

Gappa~\cite{Daumas2010} is a theorem prover for (specific forms of) VCs \Rev{about real numbers} which may involve rounding operators.
It has been shown practical for proving rounding error bounds as mentioned above~\cite{CRlibm,Boldo2011a,Boldo2013JAR}.
Gappa can be applied to our VCs but how to encode them in a way solvable with Gappa is not straightforward; 
\Rev{for example, it is unclear how to handle user-defined types (e.g. \texttt{interval}) and the special FP values $\pm\infty$ and $\mathrm{NaN}$ since Gappa does not support them natively.}



\section{Conclusion and Future Work}
\label{s:concl}

\Rev{A code for four interval \CRRev{operations}, verified with Why3,} has \RRev{been} presented.
\CRRev{We implemented the target code} with \verb|WhyML| and annotated \Rev{with properties} validity, soundness and tightness; then, \CRRev{we proved the correctness} by discharging all 
of the generated VCs by Alt-Ergo and Coq.
\RRRev{The tightness was specified in an indirect way with the \texttt{min4}/\texttt{max4} functions; we verified that each bound of a result coincides either of the extremum computed with the bounds of the inputs, or zero.}
We believe interval arithmetic code is an appealing application for program verification as its correctness matters and as the provers of this domain are developed actively.

\Rev{In the verification, several strategies were utilized, e.g., specification of auxiliary lemmas that help Alt-Ergo, and delegation of reasoning on nonlinear terms to the interactive proof with Coq.
Some lemmas are not specific to interval arithmetic; so, they can be reused by automated provers in other applications.
Many of the VCs were proved in a combined automated and interactive manner. This approach is powerful as VCs tend to become complicated, while often involving nonlinear terms. In principle, it can handle arbitrary VCs on background theories of the tools.}

As a future work, we plan to verify other interval computations, \Rev{such as} other arithmetic implementations, elementary functions, and application programs.
\Rev{We aim to schematize useful strategies for the verification of this domain.}
Another line of work is to \Rev{design and integrate} a set of axioms for interval analysis into a prover like Alt-Ergo~\cite{Conchon2017} for reasoning \Rev{about} statements on interval arithmetic.

\section*{Acknowledgements}

\RRRev{The authors thank Masahide Kashiwagi who developed the kv library and the developers of the verification tools we utilized.
We also thank Guillaume Melquiond for suggesting us the use of lemmas on reals in the early stage.
We are grateful for the many valuable comments of the peer reviewers.}
This work was partially funded by JSPS (KAKENHI \Rev{18K11240} and \Rev{18H03223}).

\bibliography{vkv}

\appendix

\section{Proof of Theorem 1}
\label{a:proof}

\RRev{
%
%
We check the tightness of the right-hand side (rhs) for the following two cases:
(i) Except when \CRRev{all} of the computations with the interval bounds become $\mathrm{NaN}$, the tightness is obvious since the bounds of the rhs are the rounded values of the operation results, i.e., the hull of some theoretical values.
(ii) For the case when all boundary computations result in $\mathrm{NaN}$, which is evaluated to $0$ with the second case of \eqref{eq:minmax4}, we confirm that it only happens in the multiplication and it results in a correct interval:
\begin{itemize}
\setlength\parskip{0em}
\setlength\itemsep{0em}
\item[$+$)] The extremum bounds are always computed as $[\RndD(\LB{x}+\LB{y}),\RndU(\UB{x}+\UB{y})]$.
Here, additions $\pm\infty + \mp\infty$ that result in $\mathrm{NaN}$ never happen;
so, the second cases of $\eqref{eq:minmax4}$ do not apply.
\item[$-$)] The rhs is always computed as $[\RndD(\LB{x}-\UB{y}),\RndU(\UB{x}-\LB{y})]$.
Neither subtractions $\pm\infty - \pm\infty$ that result in $\mathrm{NaN}$ nor
the second cases of $\eqref{eq:minmax4}$ happen.
\item[$\times$)]
    The computation results of the rhs are classified as follows:
\[
    \begin{tabular}{c||c|c|c|c}
        & $\LB{y}<0,\UB{y}\leq0$ & $\LB{y}<0,\UB{y}>0$ & $\LB{y}\geq0,\UB{y}>0$ & $\LB{y}=\UB{y}=0$ \\
        \hline
        \hline
        $\LB{x}<0,\UB{x}\leq0$ & 
        $[\RndD(\UB{x}\times\UB{y}),\RndU(\LB{x}\times\LB{y})]$ & 
        $[\RndD(\LB{x}\times\UB{y}),\RndU(\LB{x}\times\LB{y})]$ & 
        $[\RndD(\LB{x}\times\UB{y}),\RndU(\UB{x}\times\LB{y})]$ & 
        $[0,0]$ (**) \\
        \hline
        $\LB{x}<0,\UB{x}>0$ & 
        $[\RndD(\UB{x}\times\LB{y}),\RndU(\LB{x}\times\LB{y})]$ & 
        (*)
        & 
        $[\RndD(\LB{x}\times\UB{y}),\RndU(\UB{x}\times\UB{y})]$ & 
        $[0,0]$ (**) \\
        \hline
        $\LB{x}\geq0,\UB{x}>0$ &
        $[\RndD(\UB{x}\times\LB{y}),\RndU(\LB{x}\times\UB{y})]$ & 
        $[\RndD(\UB{x}\times\LB{y}),\RndU(\UB{x}\times\UB{y})]$ & 
        $[\RndD(\LB{x}\times\LB{y}),\RndU(\UB{x}\times\UB{y})]$ & 
        $[0,0]$ (**) \\
        \hline
        $\LB{x}=\UB{x}=0$ &
        $[0,0]$ (**) & $[0,0]$ (**) & $[0,0]$ (**) & $[0,0]$ \\
    \end{tabular}
\]
The cell (*) is computed as $[\min\{\RndD(\LB{x}\times\UB{y}),\RndD(\UB{x}\times\LB{y})\}, \max\{\RndU(\LB{x}\times\LB{y}),\RndU(\UB{x}\times\UB{y})\}]$.
Note that the multiplications of $0$ and $\pm\infty$ result in $\mathrm{NaN}$.
These combinations never happen in the cells not marked with (**).
For the cells marked with (**), we compute $[0,0]$ as $\forall r \in \mathbb{R}, 0\times r = 0$.
When multiplying $[0,0]$ and half-bounded intervals, $[0,0]$ is computed with the first cases of \eqref{eq:minmax4}.
When multiplying $[0,0]$ and $[-\infty,+\infty]$, $[0,0]$ is computed with the second cases.
\item[$\div$)]
    The computation results of the rhs are classified as follows:
\[
    \begin{tabular}{c||c|c}
        & $\UB{y}<0$ & $\LB{y}>0$ \\
        \hline
        \hline
        $\UB{x}<0$ & 
        $[\RndD(\UB{x}\div\LB{y}),\RndU(\LB{x}\div\UB{y})]$ & 
        $[\RndD(\LB{x}\div\LB{y}),\RndU(\UB{x}\div\UB{y})]$ \\
        \hline
        $\LB{x}\leq0,\UB{x}\geq0$ & 
        $[\RndD(\UB{x}\div\UB{y}),\RndU(\LB{x}\div\UB{y})]$ & 
        $[\RndD(\LB{x}\div\LB{y}),\RndU(\UB{x}\div\LB{y})]$ \\
        \hline
        $\LB{x}>0$ &
        $[\RndD(\UB{x}\div\UB{y}),\RndU(\LB{x}\div\LB{y})]$ & 
        $[\RndD(\LB{x}\div\UB{y}),\RndU(\UB{x}\div\LB{y})]$ \\
    \end{tabular}
\]
The divisions $\pm\infty \div \pm\infty$ that result in $\mathrm{NaN}$ never happen in all the cells;
so, the second cases do not apply.
\end{itemize}
}

\RRev{Almost identical analysis to the above is described in \cite{Kashiwagi2015}, in which the cells (**) of the multiplication table are analyzed differently.}

\section{Specification of Subtraction and Division}
\label{a:other}

The verification processes for other operations are similar to that for the multiplication. 
Some notable annotations made to the subtraction and division operators are described below.

Specification of the subtraction operator is shown in Figure~\ref{c:sub}.
In the specification of $Q_T$, the precondition \texttt{mul\char`_nan\char`_case} is omitted since it is not necessary;
if the precondition was violated, the result would be $[\mathrm{NaN},\mathrm{NaN}]$, which is an invalid interval; it is verified with the postcondition $Q_V$ that it will not happen.
At Lines~12--13, two properties on the lower bound of the result and the argument intervals are annotated;
the properties are introduced as premises in the VCs for the postconditions; 
they accelerated the proof process of Alt-Ergo in our experiment.
As a side note, similar assertions were not annotated to the addition operator, since it was verified automatically and efficiently without them.

Specification of the division operator is shown in Figure~\ref{c:div}.
Line~9 raises the \texttt{Division\char`_by\char`_zero} exception and it is verified with the postcondition $Q_Z$ at Line~3 that it is raised properly.

\begin{figure}[t]
  \lstset{frame=single}
  \lstset{numbers=left}
  \begin{lstlisting}
  let subtraction (x y: interval) : interval
    requires { valid x /\ valid y }
    ensures { valid result }
    ensures { forall u v. real_in u x /\ real_in v y -> real_in (u - v) result }
    ensures { result.inf .= min4
        (sub_down x.inf y.inf) (sub_down x.inf y.sup) 
        (sub_down x.sup y.inf) (sub_down x.sup y.sup) }
    ensures { result.sup .= max4
        (sub_up x.inf y.inf) (sub_up x.inf y.sup)
        (sub_up x.sup y.inf) (sub_up x.sup y.sup) }
  = let r = { inf = sub_down x.inf y.sup; sup = sub_up x.sup y.inf } in
    assert { r.inf .<= sub_down x.inf y.inf \/ is_nan (sub_down x.inf y.inf) };
    assert { r.inf .<= sub_down x.sup y.sup \/ is_nan (sub_down x.sup y.sup) };
    r
  \end{lstlisting}
    \caption{Code for subtraction of intervals.}
  \label{c:sub}
\vspace{1em}
  \lstset{frame=single}
  \lstset{numbers=left}
  \begin{lstlisting}
  let division (x y: interval) : interval
    (* pre- and postconditions are omitted *)
    raises { Division_by_zero -> real_in 0. y }
  = if y.inf .> zeroF then
      (* three branches are omitted *)
    else if y.sup .< zeroF then
      (* three branches are omitted *)
    else
      raise Division_by_zero
  \end{lstlisting}
    \caption{Code snippet for division of intervals: Exceptional case.}
  \label{c:div}
\end{figure}

\end{document}